# The ESO Science Archive Facility: Status, Impact, and Prospects

Martino Romaniello[1]
Magda Arnaboldi[1]
Mauro Barbieri[2]
Nausicaa Delmotte[1]
Adam Dobrzycki[1]
Nathalie Fourniol[1]
Wolfram Freudling[1]
Jorge Grave[1]
Laura Mascetti[2]
Alberto Micol[1]
Jörg Retzlaff[1]
Nicolas Rosse[2]
Tomas Tax[2]
Myha Vuong[2]
Olivier Hainaut[1]
Marina Rejkuba[1]
Michael Sterzik[1]

[1] ESO
[2] Terma GmbH

Scientific data collected at ESO's observatories are freely and openly accessible online through the ESO Science Archive Facility. In addition to the raw data straight out of the instruments, the ESO Science Archive also contains four million processed science files available for use by scientists and astronomy enthusiasts worldwide. ESO subscribes to the FAIR (Findable, Accessible, Interoperable, Reusable) guiding principles for scientific data management and stewardship. All data in the ESO Science Archive are distributed according to the terms of the Creative Commons Attribution 4.0 International licence (CC BY 4.0).

## Introduction

The science data collected at ESO's La Silla Paranal Observatory (LPO) are accessible through the ESO Science Archive Facility (SAF). The observatory comprises three sites in northern Chile's Atacama region, namely La Silla[1], Paranal[2] and the Chajnantor plateau (the Atacama Pathfinder EXperiment, or APEX telescope[3]). Data from the Atacama Large Millimeter/submillimeter Array (ALMA) observatory[4] are also directly accessible from the ESO Science Archive, so that they can be conveniently queried together with the data from LPO. In addition, ESO also hosts and operates the European copy of the dedicated ALMA Science Archive[5], which provides extended search capabilities tailored to these data; it was recently described by Stoehr et al. (2022).

At the time of writing, the ESO Science Archive contains, in a uniform and consistent form, data from more than 30 instruments (and counting), covering a wide range of observing techniques, data types and formats, and their metadata. It stores all the raw science data and the related calibrations. A growing selection of processed data is also available, on which science measurements can be readily performed. The archive home page is at: archive.eso.org. User support and a knowledgebase database are provided at support.eso.org.

ESO has a long tradition of fostering 'Open Access' to scientific data, and it endorses the European EOSC initiative[6]. As an overarching principle, ESO subscribes to the FAIR[7] (Findable, Accessible, Interoperable, Reusable) guiding principles for scientific data management and stewardship (Wilkinson et al., 2016). Access to the ESO Science Archive is regulated by policy[8]. In general terms, the Principal Investigators (PIs) of successful proposals for observing time on ESO telescopes, along with their delegates, have exclusive access to their scientific data for a proprietary period, after which the data are accessible to all users in the worldwide community. The default proprietary period is set by the ESO Director General and communicated at the time of proposing for observing time[9]. It is typically one year, but may depend on the observing programme type, as detailed in the policy (for example, Public Survey raw data are immediately public). Processed data distributed via the ESO Science Archive retain the same proprietary protections as the raw data they were derived from. All data in the ESO archive retain ESO's copyright and are distributed according to the terms of the Creative Commons Attribution 4.0 International licence (CC BY 4.0[10]). The use of ESO data, for example in publications, either downloaded directly from the ESO Science Archive or via third parties, must be acknowledged[8].

## Data content and access

Over the course of the last 25 years, a large fraction of the accessible sky has been observed by ESO telescopes. The density map coverage of the science raw data available in the ESO Science Archive as of June 2023 is shown in Figure 1. Given the wide variety of data distributed over such a large area, it may be a challenge to find the data needed. The Science Archive provides several means to do so, tailored to the different use cases.

The raw data can be queried by basic instrumental, target, observing programme and scheduling criteria through a unified form[11]. Specialised query forms for individual instruments, which expose many more detailed technical and scientific search parameters, are also available[12]. Once a user has selected the raw science data of interest, relevant calibration files can be associated automatically prior to download. The service is tuned to provide the calibrations as defined in the instrument's Calibration Plan[13]. At this stage, users can choose whether they want raw and/or pre-processed master calibrations. They can also select to include night-log information, such as weather conditions and notes from the observer. Once downloaded, users can process raw data along with the associated calibrations to remove signatures from the telescope, instrument and Earth's atmosphere, and to calibrate the resulting data products in physical units. For this, dedicated software tools to process and organise the data and the execution sequence are made available[14]. At this point, the data are ready for extraction of the science signal and its subsequent analysis.

The current era in astronomy research is characterised by an abundance of data and the need to combine them across facilities, wavelengths, and messengers. It is, therefore, imperative to lower as much as possible the user's access barrier to the data. The goal is to reach as wide an audience as possible, providing a complete overview of the content of the archive, while requiring as little overhead as possible on the part of the researcher. To this end, the ESO Science Archive provides access to processed data. Via this route, users can download data that have already gone through most of the processing needed in preparation for extracting the science signal and are thus free from atmospheric and instrumental effects and are calibrated in physical units. The main science files are accompanied and complemented by ancillary ones that provide additional information useful to their exploitation (for example, 2D calibrated spectra are often provided with the main 1D product to allow a custom spectrum extraction; and white-light images go with data cubes). Each data collection comes with extensive textual documentation in the form of a Release Description. In most cases, processed data from the archive are directly ready for science analysis.

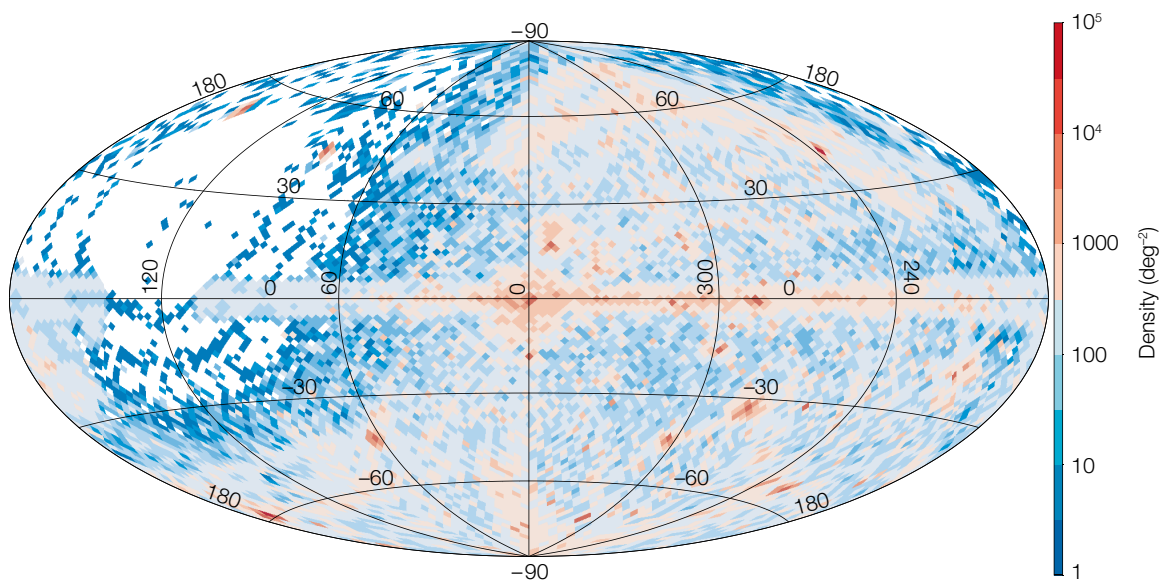

Figure 1. Sky density map of the raw data content of the ESO Science Archive Facility as of June 2023. Popular locations on the sky, such as the Galactic centre or the Magellanic System, clearly stand out, as do the footprints of some of the Public Surveys. The few observations close to the Celestial North Pole are clearly the result of errors in metadata content.

There are two main sources of such processed data. One is provided by users who carried out the processing, typically, but not always, for their own projects, and returned the results to the SAF. This is mandatory for observing programmes that require large, coordinated amounts of telescope time, namely Public Surveys[15] and Large Programmes[16], as well as for Hosted Telescopes where there is a signed agreement with ESO for this. In these cases, generating data, both raw and processed, with a long-lasting legacy value is an important criterion in the programme selection process. Voluntary contributions from individual users are much encouraged; this provides a great way to give data, and their authors, enhanced visibility and citeability. To this end, each data collection in the ESO Science Archive is assigned a unique persistent Digital Object Identifier[17] (DOI). Its processing is tailored to the science case(s) of the originating observing programmes, and results are often described and used by the team in refereed publications. Typical data products include calibrated deep and/or mosaicked images and data cubes, stacked spectra, and flux maps. In several cases, these are used to generate source catalogues. These are the highest-level processed data and contain directly the physical parameters of the celestial sources.

The other main channel of influx of processed data for the Science Archive is carried out at ESO. Here, the data histories of instruments, or instrument modes, are processed as consistently and as completely as possible and ingested into the SAF. By its very nature, this data processing is not tailored to any specific science case, but is focused on removing the instrumental and atmospheric signatures and on calibrating in physical units large, coherent datasets. The tools used in-house to process the data are the same ones that are made publicly available[14]. The impact of archival processed data is discussed further below.

Stewardship of science data product: the Phase 3 process

At the time of writing, the ESO Science Archive contains four million processed science files from nearly 80 data collections, covering virtually all data types and observational techniques enabled by the slew of more than 30 instruments that ESO operates at LPO. They cover a correspondingly large range of observing techniques, data types, formats, and metadata.

Without science-oriented data stewardship, curation and homogenisation, the archive would be just a big bucket of bits and bytes, where finding data would be exceedingly hard and reserved to a few experts. Therefore, before ingestion into the archive, the processed data undergo an auditing process for completeness, compliance, consistency, and documentation[18] (Arnaboldi et al., 2011). This process is a collaborative effort between the data provider and ESO and is called Phase 3, reflecting the fact that it closes the loop after the solicitation and handling of observing proposals (Phase 1) and the observation preparation and execution at the telescope (Phase 2).

To ensure data consistency and accessibility throughout this broad variety of archive holdings, the Phase 3 process enforces the use of the ESO Science Data Product Standard[19], an interface document that defines the data format and metadata (content and definition) for the various types (images, spectra, cubes, interferometric visibilities, catalogues, and so on). It specifies how to encode the level of calibration, scientific quality, originating files (provenance), ancillary data and product versioning. Its content is continuously evolving to reflect the evolving data landscape, for example to incorporate new data types produced by different observing techniques and facilities. The ESO Science Data Product Standard incorporates accepted Virtual Observatory (VO) standards, making ESO data interoperable with the other VO resources. Compliance with the standard is a fundamental requirement which ensures that data can be easily located among the broader ESO holdings and in the general global data scenario. This is a must in the current era of multi-instrument, multi-wavelength, multi-messenger astronomy.

### The Archive Science Portal

An important driver for the metadata curation and homogenisation ensuing from the Phase 3 process is that data can be presented and queried uniformly across collections, independently of their origins and specificities. We have built different ways to browse the processed data in the SAF, namely web interfaces, and programmatic and scripted access.

The web interfaces offer a low-barrier access by presenting the data and metadata in an intuitive graphical interface[20]. Query parameters are represented by elements in the page that have the dual function of visually expressing the content of the archive and rendering the user's choices. The results are then rendered on the backdrop of the celestial sphere and included in a tabular form, which summarises their main characteristics (see Figure 2, top panel). Given the underlying compatibility with VO protocols, the results can be sent to VO-aware tools, such as, for example, TOPCAT or Aladin.

Figure 2. Top panel: The landing page of the web interface to the ESO Archive Science Portal[20]. Bottom panel: Example of the web page where individual datasets can be explored in detail. In this case, the file is a MUSE datacube.

Upon request, individual datasets can be explored in detail through previews that are customised by data type. As an example the preview of a Multi Unit Spectroscopic Explorer (MUSE) datacube is shown in the right panel of Figure 2.

A dedicated web interface is available to query source catalogue data[21]. Once users select the catalogue they are interested in, they can constrain the search on any combination of its columns.

Repetitive or otherwise particularly demanding tasks can be coded and automatised by using the provided programmatic access[22]. It too makes extensive use of VO protocols, thus ensuring interoperability and a high level of standardisation. Extensive documentation and example ADQL and TAP queries, Python scripts and Jupyter notebooks are provided to guide users in their first steps, and throughout some more complex cases.

A forum platform is available for archive researchers to exchange ideas and questions[23].

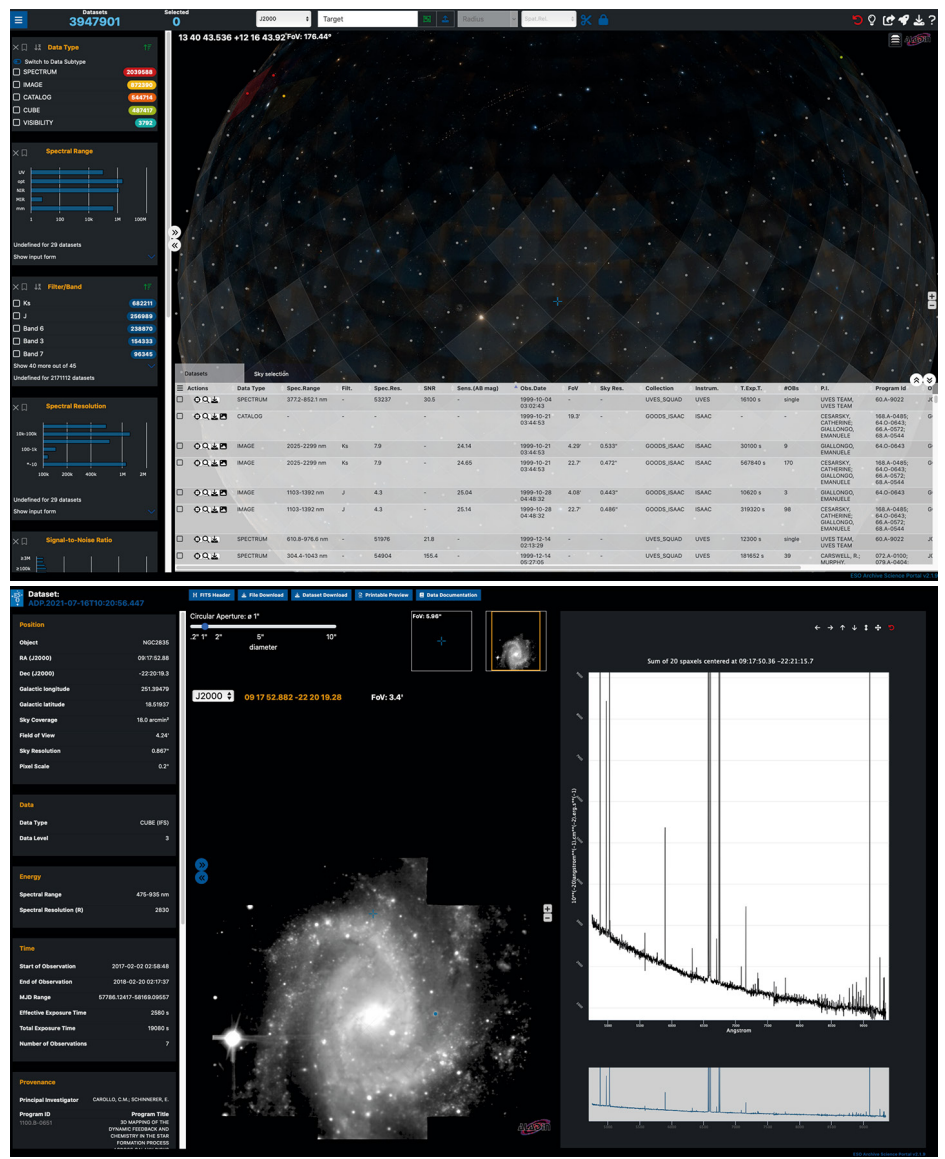

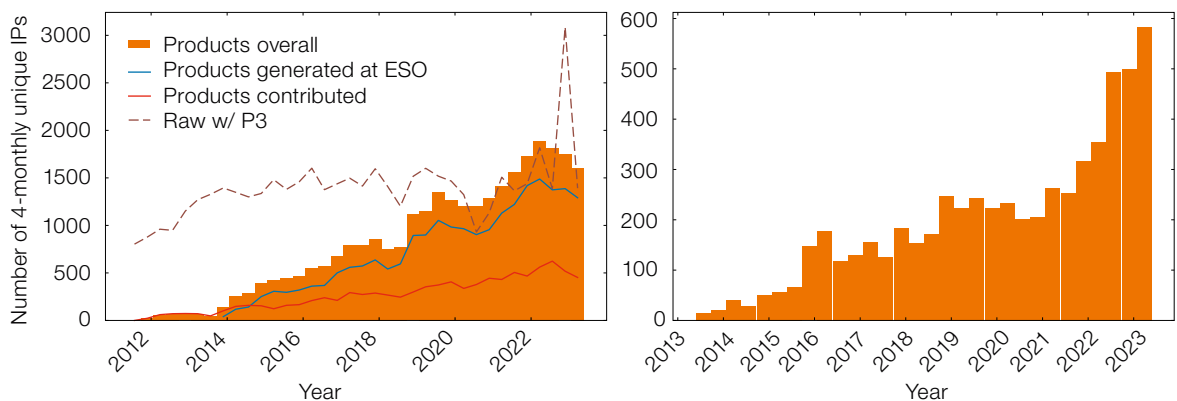

Figure 3. The differential number of unique IP addresses as a function of time from which processed files (left panel) and source catalogues (right panel) in the ESO Science Archive are accessed. The IP addresses are a proxy for the number of users, with each using on average 1.5 IPs. Resorting to IP addresses as a proxy for users is made necessary by the fact that the vast majority of downloads are anonymous.

## The impact of the ESO Science Archive Facility

All of the assets in the ESO SAF, i.e., raw data and products generated at ESO or contributed by the community and catalogues, are in great and increasing demand. This is shown in Figure 3, where the number of unique IP addresses, a proxy for the users downloading data, is plotted as a function of time for processed files and source catalogues (left and right panel, respectively). Interestingly, the increase in the download of processed data did not come at the expense of the access to raw data, just as the fast increase in the number of users of data processed by ESO has not hindered the need for data generated externally (bottom panel in Figure 3). The different types of data are, then, highly complementary.

Figure 4 shows the contribution of the SAF to the science output of LPO. This is expressed in terms of the fraction of refereed papers using LPO data that made use of the archive (a referred paper is classified by the ESO Library as archival if there is no overlap between its authors and the members of the original observing proposal[24]). There is a clear upward trend since the inception of the ESO Science Archive in its current form at the end of the 1990s. Currently, about 4 papers out of 10 utilising LPO data make use of the ESO Science Archive.

The ESO Science Archives, both LPO and ALMA, featured as a Special Topic at the 47th meeting of the ESO Users Committee, which was held on 20 and 21 April 2023[25]. The high level of user satisfaction with the archives was confirmed in the discussions during the meeting, as well as in the Committee's report[26], which is based on a poll of the science community.

## What's next

As discussed above, the availability of processed data has led to a tangible boost to the access and usefulness of the ESO Science Archive. The engagement of the community at large in providing reduced data has been very successful, and the archive now provides more than 60 out of 80 collections secured in this way. This number is, of course, poised to increase as the policy mandating the delivery of processed data for new Public Surveys, Large Programmes and Hosted Telescopes/Instruments continues and will also include data from ESO's Extremely Large Telescope (ELT) in the future.

In addition to maintaining support for the Phase 3 process for data provided by internal and external users, we are also exploring new ways of collaborating with the community that are not linked to specific observing programmes. Prominent examples include the data stream for the Precision Integrated-Optics Near-infrared Imaging ExpeRiment (PIONIER; Le Bouquin et al., 2011), and the VISTA EXtension to Auxiliary Surveys (VEXAS; Spiniello & Agnello, 2019) and Ultraviolet and Visual Echelle Spectrograph Spectral QUasar Absorption Database (UVES SQUAD; Murphy et al., 2019) collections. We have established a collaboration with the High Contrast Data Center[27] (HC-DC, previously the SPHERE Data Center) in Grenoble. Data from the Spectro-Polarimetric High-contrast Exoplanet REsearch instrument (SPHERE) are regularly processed there, leveraging the considerable science expertise available, and delivered to ESO for wide dissemination (a public archive copy is also maintained at the HC-DC). The first data products were published in December 2022 in the ESO Science Archive Facility and include imaging data from the InfraRed Dual Imaging and Spectrograph (IRDIS) subsystem, observed during the ESO Periods 103 and 104, i.e., acquired between April 2019 and March 2020. Both the time and the instrumental modes covered will expand in future releases. Similarly, a new collaboration is being set up with the Very Large Telescope Interferometer (VLTI) Expertise Centres of the OPTICON Radionet Pilot[28]. In this case, the processing of GRAVITY data up to calibrated visibilities is performed at ESO, while the Expertise Centres provide scientific guidance and quality control.

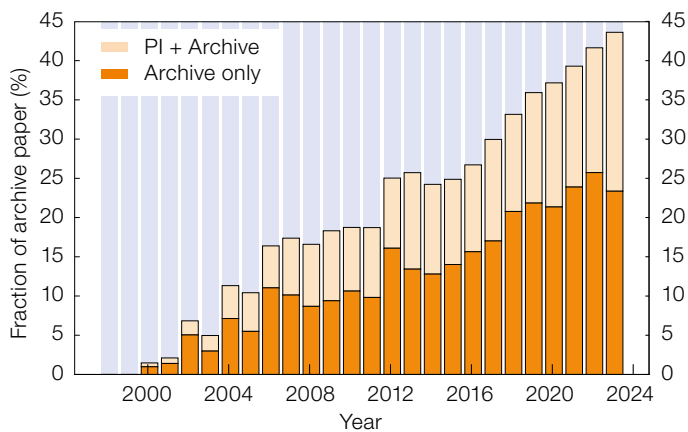

Figure 4. The fraction of refereed publications using La Silla Paranal data that made use of the archive, entirely (dark orange bars) or partially (light yellow bars). Source: ESO Telescope Bibliography[24].

The evolution of data generated internally at ESO for publication in the archive will be along two main directions. Firstly, we are working towards making the processed data for the new instruments available much sooner than was possible in the past. The goal is to do so at the same time as the raw data first become public, typically about a year after the start of science operations. This initial delay is determined by the need to characterise the data calibration and processing well enough to provide products of known and documented quality and accuracy. And secondly, to increase the quality of the data processed at ESO we are implementing a more extensive in-depth quality control aimed at identifying ways to improve the products. This is complementing the reprocessing of entire data streams in case of significant improvement of the pipeline and/or calibrations.

With the data content increasing in quality, quantity and complexity, the archive tools to browse and access them must evolve too. As an example, the spectroscopic surveys with the 4-metre Multi-Object Spectrograph Telescope (4MOST) alone will return each year more than three times as many individual processed files as we have collected in the last ten years. The main drivers for this evolution are towards a unification of the web query interfaces to the data, and towards querying the archive content by (selected) physical properties of the astronomical sources. The former aims to reduce the complexity for users by providing as far as possible a single experience where currently different interfaces are in place (for example for processed data[20] and source catalogues[21]). With the latter, we instead aim to provide query capabilities that are closer to the science questions that archive users have. In both cases, the overarching objective is to help scientists to get quickly, efficiently and accurately to the data of interest among the millions of assets that are stored and preserved in the treasure trove which is the ESO Science Archive.

### Acknowledgements

The list of people who have made possible the growth and success of the ESO Science Archive Facility goes far beyond the authors of this article. We would like to extend our thanks to the ESO colleagues who have worked with us throughout the years, especially the software development and testing team (Vincenzo Forchì, Ahmed Mubashir Khan, Uwe Lange, Stanislaw Podgorski, Fabio Sogni, Malgorzata Stellert, and Stefano Zampieri). The work, time and dedication of colleagues from the scientific community who have provided processed data to the ESO Science Archive is gratefully acknowledged: their contributions represent a truly invaluable science resource. This article is dedicated to the memory of our colleague Jörg Retzlaff. His hard work, dedication and talent were fundamental in making the ESO Science Archive Facility the powerful science resource for the whole community that it is today. It was an honour and a pleasure to work with him, he will be fondly remembered.

### Links

1. ESO La Silla: https://www.eso.org/public/teles-instr/lasilla
2. ESO Paranal: https://www.eso.org/public/teles-instr/paranal-observatory
3. APEX: https://www.eso.org/public/teles-instr/apex
4. ESO ALMA webpage: https://almascience.eso.org
5. ESO ALMA Science Archive: https://almascience.eso.org/aq
6. EOSC website: https://eosc.eu
7. FAIR principles: https://www.go-fair.org/fair-principles
8. ESO Science Archive data access policy: http://archive.eso.org/cms/eso-data-access-policy.html
9. ESO Phase 1 Call for Proposals: https://www.eso.org/sci/observing/phase1.html
10. CC BY 4.0 attribution: https://creativecommons.org/licenses/by/4.0
11. ESO raw data query form: http://archive.eso.org/eso/eso_archive_main.html
12. ESO instrument-specific query forms: http://archive.eso.org/cms/eso-data/instrument-specific-query-forms.html
13. ESO instrument calibration plan: https://www.eso.org/sci/observing/phase2/SMGuidelines/CalibrationPlan.generic.html
14. ESO VLT instrument pipelines: https://www.eso.org/sci/software/pipelines
15. ESO Public Surveys: https://www.eso.org/sci/observing/PublicSurveys/sciencePublicSurveys.html
16. ESO Large Programmes: https://www.eso.org/sci/observing/teles-alloc/lp.html
17. ESO Science Archive DOIs: https://archive.eso.org/wdb/wdb/doi/collections/query
18. ESO Phase 3 process: https://www.eso.org/sci/observing/phase3.html
19. ESO Science Data Products Standard: http://www.eso.org/sci/observing/phase3/p3sdpstd.pdf
20. ESO Archive Science Portal: http://archive.eso.org/scienceportal
21. ESO catalogue query: https://www.eso.org/qi
22. ESO Science Archive programmatic access: http://archive.eso.org/programmatic
23. ESO archive community forum: esocommunity.userecho.com
24. ESO Telescope Bibliography: https://telbib.eso.org
25. ESO Users Committee 47th meeting: https://www.eso.org/public/about-eso/committees/uc/uc-47th.html
26. ESO Users Committee report of 47th meeting: https://www.eso.org/public/about-eso/committees/uc/uc-47th/UC47_2023_UCreport.pdf
27. High Contrast Data Center: https://sphere.osug.fr/spip.php?rubrique16&lang=en
28. OPTICON RadioNet Pilot: https://www.orp-h2020.eu